# On Diósi-Penrose criterion of gravity-induced quantum collapse

Shan Gao
Unit for HPS & Centre for Time, SOPHI, University of Sydney
Email: sgao7319@uni.sydney.edu.au

It is shown that the Diósi-Penrose criterion of gravity-induced quantum collapse may be inconsistent with the discreteness of space-time, which is generally considered as an indispensable element in a complete theory of quantum gravity. Moreover, the analysis also suggests that the discreteness of space-time may result in rapider collapse of the superposition of energy eigenstates than required by the Diósi-Penrose criterion.

The physical origin of quantum collapse is a fundamental problem of quantum theory. It has been conjectured that the wavefunction collapse may be caused by gravity. The gravity-induced quantum collapse conjecture can be traced back to Feynman [1]. In his *Lectures on Gravitation*, he considers the problems of quantizing macroscopic objects and contemplates on a possible breakdown of quantum theory due to gravity. The idea has been developed by many authors since then (e.g. [2-5]), among them Diósi and Penrose presented a very specific argument and model [4-5]. They argued that when a superposition of two spatially displaced states of the same object will reduce to either one state or the other, although from different physical considerations. Moreover, they independently obtained the same criterion of gravity-induced quantum collapse, which is usually called the Diósi-Penrose criterion. Since Penrose's argument is minimalist in conception, which relies only on the fundamental principles of quantum mechanics and general relativity, we will mainly discuss his proposal in this Letter.

In Penrose's gravity-induced collapse model [5], the collapse time formula is

$$\tau_c \approx \frac{\hbar}{\Delta E_G} \qquad (1)$$

where $\hbar$ is Planck constant divided by $2\pi$, and $\Delta E_G$ is the gravitational self-energy of the difference between the mass distributions belonging to the two states in the superposition

$$\Delta E_G = \frac{1}{G}\int (\nabla \Phi_2 - \nabla \Phi_1)^2 dx^3 \qquad (2)$$

where $\Phi_1$ and $\Phi_2$ are the Newtonian gravitational potentials of the two states, and $G$ is Newton's gravitational constant. Although the Diósi-Penrose criterion denoted by Eq. (1) seems consistent with existing experiments and macroscopic experience, we will show that at least for some situations it may contradict the requirement of the discreteness of space and time, which is generally considered as an indispensable element in a complete theory of quantum gravity. This can be seen from the following analysis of a typical example.

Consider a quantum superposition of two different energy eigenstates. Each eigenstate has a well-defined static mass distribution in the same spatial region with radius $R$. For example, they are rigid balls of radius $R$ with different uniform mass density. The initial state is

$$\psi(x,0) = \frac{1}{\sqrt{2}}[\varphi_1(x) + \varphi_2(x)] \qquad (3)$$

where $\varphi_1(x)$ and $\varphi_2(x)$ are two energy eigenstates with energy eigenvalues $E_1$ and $E_2$ respectively. According to the Diósi-Penrose criterion, we have

$$\Delta E_G \approx \frac{G(\Delta E)^2}{c^4 R} \qquad (4)$$

$$\tau_c \approx \frac{\hbar}{\Delta E_G} \approx \frac{c^4 \hbar R}{G(\Delta E)^2} \qquad (5)$$

where $\Delta E = E_2 - E_1$ is the energy difference, $c$ is the speed of light. This means that after the collapse time $\tau_c$ the superposition will be greatly destroyed, but long before this time the superposition should be very precise. Then according to the linear Schrödinger evolution, we have:

$$\psi(x,t) = \frac{1}{\sqrt{2}}[e^{-iE_1 t/\hbar}\varphi_1(x) + e^{-iE_2 t/\hbar}\varphi_2(x)] \qquad (6)$$

and

$$\rho(x,t) = |\psi(x,t)|^2 = \frac{1}{2}[\varphi_1^2(x) + \varphi_2^2(x) + 2\varphi_1(x)\varphi_2(x)\cos(\Delta E t/\hbar)] \qquad (7)$$

This indicates that the probability density $\rho(x,t)$ will oscillate with a period $T = h/\Delta E$ in each position of space before the collapse time $\tau_c$. Since the radius $R$ is much larger than the Planck length $L_P = (\hbar G/c^3)^{1/2}$ in practical situations, even when the energy difference $\Delta E$ reaches the Planck energy $E_P = (\hbar c^5/G)^{1/2}$, the oscillation will always happen, as the collapse time of the superposition

$$\tau_c \approx \frac{c^4 \hbar R}{G(\Delta E)^2} = \frac{R}{c} \qquad (8)$$

is still much greater than the period of oscillation $T = h/E_P \approx L_P/c \equiv T_P$, where $T_P$ is the Planck time. Then when the energy difference $\Delta E$ exceeds the Planck energy $E_P$ such as $\Delta E > 2\pi E_P$, the probability density $\rho(x,t)$ will oscillate with a period shorter than the Planck time $T_P$. However, the Planck time $T_P$ is generally assumed as the minimum distinguishable size of time in discrete space-time (see, e.g. [6-9]), and no physical change can happen during a time interval shorter than it. Thus we reach a contradiction between the Diósi-Penrose criterion of quantum collapse and the requirement of discrete space-time[1].

---

[1] Here it is implicitly assumed that the wavefunction collapse is mainly due to Penrose's mechanism and predicted by the Diósi-Penrose criterion.

It has been widely argued that the proper combination of quantum theory and general relativity, two firm results of which are the formula of black hole entropy and the generalized uncertainty principle [6-9], may inevitably result in the discreteness of space and time. Moreover, the arguments show that the minimum time interval and the minimum length are respectively of the order of Planck time and Planck length in discrete space and time. For example, the minimum length can be derived from the following generalized uncertainty principle [7-8]:

$$\Delta x = \Delta x_{QM} + \Delta x_{GR} \geq \frac{\hbar}{2\Delta p} + \frac{2L_p^2 \Delta p}{\hbar} \qquad (9)$$

Since a full description of quantum gravity in terms of discrete space-time is not yet available, and the formulations and meanings of discrete space-time are different in the existing theories (*e.g.* string theory, loop quantum gravity, and quantum geometry etc), here we only resort to a minimum explanation of the discreteness of time, namely that a time interval shorter than the shortest time is physically meaningless, and it cannot be measured in principle either. As a result, any physical process can only happen during a time interval not shorter than the minimum time interval.

In order to avoid the contradiction with the discreteness of space and time, the quantum collapse should happen more early than the Diósi-Penrose criterion requires for the above superposition. Concretely speaking, when the energy difference $\Delta E$ reaches the Planck energy $E_P$, the superposition should collapse into one of its branches during a Planck time scale $T_P$, so that the probability density $\rho(x,t)$ will not oscillate with a period shorter than this minimum time scale. Therefore, the discreteness of space and time will result in a rapider quantum collapse for the above superposition than required by the Diósi-Penrose criterion. It may be worth noting here that this restriction of discrete space and time favors the energy-driven collapse models (e.g. [10-13]), according to which the collapse time formula is

$$\tau_c \approx \frac{\hbar E_P}{(\Delta E)^2} \qquad (10)$$

Indeed, this formula requires that when the energy difference $\Delta E$ is about the Planck energy $E_P$, the collapse time is about the Planck time $T_P$. However, as rightly pointed out by Pearle [14], the energy-driven collapse models cannot consistently account for the existing experiments, as well as the definiteness of macroscopic objects. In fact, in the above specific example, the energy difference is also equivalent to some kind of difference of space-times [15]. Therefore, it is still consistent with Penrose's viewpoint.

As we think, Penrose's trenchant argument for the gravity-induced quantum collapse, based on the deep and profound incompatibility between the principle of superposition of quantum mechanics and the principle of general covariance of general relativity, is still valid, only his collapse criterion may be not right. This is not surprising, since, as Penrose had already stressed [5], although there does exist an essential ill-definedness of the notion of time-translation (or uncertainty in the time-translation Killing vector) for a quantum superposition of different space-times, the uncertainty formula Eq. (2) proposed by him is provisional. First, the uncertainty in the time-translation Killing vector for the superposed space-times is a "velocity uncertainty", while the uncertainty denoted by Eq. (2) is an "acceleration uncertainty", and their precise relationship is model-dependent. Next, Eq. (2) does not consider the uncertainty in the identification of the actual time coordinate for the superposed space-times, which may be very important in full general relativity. In some sense, the above requirement of the discreteness of time may relate to this uncertainty [15]. Moreover, the direct "position uncertainty" may be also relevant, although it does not appear in the uncertainty in the time-translation Killing vector. Thirdly, the collapse time formula denoted by Eq. (1) seemingly has no firm physical basis when carefully examined. The argument resorting to the application of Heisenberg's uncertainty principle to unstable particles or states only has a very limited force, as these two situations are obviously different in nature; one is within standard quantum mechanics, while the other is already beyond it. In particular, contrary to the gravity-induced quantum collapse, the decay of an unstable state (e.g. excited state of atom) is actually not spontaneous but caused by the background field constantly interacting with it. In some extreme situations, the state may not decay at all when being in a very special background field with bandgap [16]. Lastly, we note that the Diósi-Penrose criterion is not right in the strictly Newtonian regime [17][2], and its microscopic formulation is unclear and still has some problems (e.g. the cut-off difficulty) [4].

To sum up, the physical origin of quantum collapse is still a great puzzle in the existing quantum theory. Penrose's observation that the incompatibility between quantum mechanics and general relativity may result in quantum collapse is surely a promising beginning to solve this puzzle. Yet his argument still needs to be further refined and developed. In particular, the precise formulation of the difference of space-times in a quantum superposition [15, 17, 18], which results in the gravity-induced quantum collapse, still needs to be found. Our analysis suggests that the formulation may relate to the discreteness of space-time. Moreover, the analysis also implies that the discreteness of space-time may result in a rapider collapse of the superposition of energy eigenstates than required by the Diósi-Penrose criterion.

**References**

[1] R. Feynman. *Feynman Lectures on Gravitation*. B. Hatfield (ed.), Reading, Massachusetts: Addison-Wesley (1995).
[2] F. Károlyházy, *Nuovo Cimento A* 42, 390-402 (1966).
[3] F. Károlyházy, A. Frenkel, and B. Lukács, On the possible role of gravity on the reduction of the wavefunction, in R. Penrose and C. J. Isham (eds.), *Quantum Concepts in Space and Time*. Oxford, England: Clarendon Press, pp. 109-128. (1986).
[4] L. Diósi, *Phys. Lett. A* 120, 377 (1987); L. Diósi, *Phys. Rev. A* 40 , 1165 (1989); L. Diósi, *Braz. J. Phys.* 35, 260 (2005); L. Diósi, *J. Phys. A: Math. Theor.* 40, 2989 (2007).
[5] R. Penrose, *Gen. Rel. Grav.* 28, 581 (1996); R. Penrose, *Phil. Trans. R. Soc. Lond. A* 356, 1927 (1998); R. Penrose, Wavefunction collapse as a real gravitational effect. In: *Mathematical Physics 2000* ed. A

---

[2] Since the Diósi-Penrose criterion, independent of the speed of light, should remain valid in the nonrelativistic Newton

Fokas *et al*. (London: Imperial College) 266–282.

[6] H. Salecker and E. P. Wigner. *Physical Review* 109, 571 (1958).

[7] L. J. Garay. *International Journal of Modern Physics A* 10, 145 (1995).

[8] R. J. Adler and D. I. Santiago. *Modern Physics Letters A* 14, 1371 (1999).

[9] L. Smolin. *Three Roads to Quantum Gravity*. Oxford: Oxford University Press (2000).

[10] I. C. Percival, *Proc. Roy. Soc. A* 451, 503 (1995) and *Quantum State Diffusion*, Cambridge: Cambridge University Press (1998).

[11] L. P. Hughston, *Proc. Roy. Soc. A* 452, 953 (1996).

[12] S. L. Adler and L. P. Horwitz, *Jour. Math. Phys.* 41, 2485 (2000).

[13] S. L. Adler, *J. Phys. A: Math. Gen.* 35, 841-858 (2002); *Phys. Rev. D* 67, 25007 (2003).

[14] P. Pearle. *Phys. Rev. A* 69, 42106 (2004).

[15] S. Gao, *Int. J. Theor. Phys.*, 45(10), 1943 (2006).

[16] E. Yablonovitch, *Phys. Rev. Lett.* 58, 2059 (1987).

[17] J. Christian, Why the quantum must yield to gravity. In: *Physics Meets Philosophy at the Planck Scale*, ed. C. Callender and N. Huggett. Cambridge: Cambridge University Press (2001) p.305.

[18] J. S. Anandan, Quantum measurement problem and the gravitational field, In *The Geometric Universe: Science, Geometry, and the Work of Roger Penrose*. ed. S. A. Huggett *et al*. Oxford: Oxford University Press (1998) pp. 357-368.

domain, this result may indicate that the criterion is already inconsistent with macroscopic experience.